\documentclass[11pt, a4paper]{article}
\usepackage[centering]{geometry}
\usepackage[english]{babel}
\usepackage{amsmath,amsfonts,amssymb,graphicx,amsthm}

\usepackage[mathlines]{lineno}
\usepackage{makeidx}
\usepackage{cite}
\usepackage{chngcntr}

\graphicspath{{./fig/}}
\newcommand{\eps}{\varepsilon}
\newcommand{\N}{\mathbb{N}}

\newcommand{\etal}{\emph{et al.}\xspace}

\usepackage[ruled,noend,linesnumbered,algosection]{algorithm2e}
\newenvironment{alg}{
  \begin{algorithm}[htbp]
    \DontPrintSemicolon
    \SetKwInput{KwIn}{input}
    \SetKwInput{KwOut}{output}
  }{\end{algorithm}}
   
\newtheorem{theorem1}{Theorem}{\bfseries}{\itshape}
\counterwithin{theorem1}{section}
\newtheorem{lem}[theorem1]{Lemma}{\bfseries}{\itshape}
\newtheorem{thm}[theorem1]{Theorem}{\bfseries}{\itshape}
\newtheorem{obs}[theorem1]{Observation}{\bfseries}{\itshape}
\newtheorem{cor}[theorem1]{Corollary}{\bfseries}{\itshape}

\title{A Time-Space Trade-off for Computing the $k$-Visibility 
Region of a Point in a Polygon\footnote{
A preliminary version appeared as 
Y.~Bahoo, B.~Banyassady, P.~Bose, S.~Durocher, and 
W.~Mulzer. \emph{Time-Space Trade-off for Finding 
the $k$-Visibility Region of a Point in a Polygon}. Proc.~11th WALCOM, 
2017.  This work was partially 
supported by DFG project MU/3501-2, ERC StG 757609, and by the 
Natural Sciences and Engineering Research Council of Canada (NSERC).}}

\author{Yeganeh Bahoo\thanks{Department of Computer Science,
        University of Manitoba, \{bahoo, durocher\}@cs.umanitoba.ca}
        \and
        Bahareh Banyassady\thanks{Institut f\"ur Informatik, Freie 
	Universit\"at Berlin,
        {\{bahareh, mulzer\}@inf.fu-berlin.de}}
        \and
        Prosenjit K. Bose\thanks{School of Computer Science,
        Carleton University, {jit@scs.carleton.ca}}
        \and 
        Stephane Durocher$^\dagger$
        \and 
        Wolfgang Mulzer$^\ddagger$
        }
\date{}
\begin{document}
\maketitle

\begin{abstract}
Let $P$ be a simple polygon with $n$ vertices, and let 
$q \in P$ be a point in $P$. Let $k \in \{0, \dots, n - 1\}$. 
A point $p \in P$ is \emph{$k$-visible} from $q$ 
if and only if the line segment $pq$ crosses the 
boundary of $P$ at most $k$ times. The \emph{$k$-visibility 
region} of $q$ in $P$ is the set of all points that 
are $k$-visible from $q$.  We study the problem of 
computing the $k$-visibility region in the 
limited workspace model, where the input resides 
in a random-access read-only memory of $O(n)$ words, 
each with $\Omega(\log{n})$ bits. The algorithm can read 
and write $O(s)$ additional words of workspace, 
where $s \in \N$ is a parameter of the model. The 
output is written to a write-only stream. 

Given a simple polygon $P$ with $n$ vertices and 
a point $q \in P$, we present an
algorithm that reports the $k$-visibility region 
of $q$ in $P$ in $O(cn/s+c\log{s} + 
\min\{\lceil k/s \rceil n,n \log{\log_s{n}}\})$ 
expected time using $O(s)$ words of workspace. 
Here, $c \in \{1, \dots, n\}$ is the number of 
\emph{critical vertices} of $P$ for $q$ where the 
$k$-visibility region of $q$ may change. We 
generalize this result for polygons with holes 
and for sets of non-crossing line segments.\\
\linebreak\textbf{Keywords:} Limited workspace model, 
$k$-visibility region, 
Time-space trade-off
\end{abstract}

\section{Introduction}
Memory constraints on mobile devices and 
distributed sensors have led to an increasing 
focus on algorithms that use their memory 
efficiently. One common approach to capture 
this notion is the \emph{limited workspace model}~\cite{asano2013memory}. 
Here, the input is provided in a random-access read-only 
array of $O(n)$ words. Each word has 
$\Omega(\log{n})$ bits. Additionally, there is a 
read/write memory with $O(s)$ words, where 
$s \in \{1, \dots, n \}$ is a parameter of the model. 
This is called the \emph{workspace} of the 
algorithm. The output is written to a write-only 
stream. 

Let $P$ be a simple polygon with $n$ vertices and $n$ edges,
and let $q$ be a point in $P$. Let
$k \in \{0, \dots, n - 1\}$.
A point $p \in P$ is \emph{$k$-visible} from $q$ 
if and only if the line segment $pq$ has at most 
$k$ proper intersections with the boundary 
$\partial P$ of $P$ ($p$ and $q$ do not count toward
the number of intersections).\footnote{For $k = n - 1$, the
whole polygon is $k$-visible from $q$, so there is no
reason to consider $k > n - 1$.} The set of $k$-visible 
points in $P$ from $q$ is called the 
\emph{$k$-visibility region} of $q$ in $P$;
see Figure~\ref{fig:fig1}.
We denote it by $V_ k(P,q)$. For $k = 0$, this 
notion corresponds to classic visibility in polygons. 

Visibility problems have played a major role 
in computational geometry since the very beginning
of the field. 
Thus, there is a rich history of previous results;
see the book by Ghosh~\cite{ghosh2007visibility} for 
an overview. The concept of $1$-visibility first appeared 
in a work by Dean~\etal~\cite{dean1988recognizing} as far
back as 1988. In the related 
\emph{superman problem}~\cite{MouawadSh94}, we are
given two polygons $P$ and $G$ such that
$G \subseteq P$, and a point $p \in P \setminus G$. 
The goal is to find the minimum
number of edges in $P$ that need to be made opaque in
order to make $G$ invisible from $p$. 
More general $k$-visibility, for $k > 1$,
is more recent. Since 2009, this variant of visibility has been
explored more widely due to its relevance in wireless networks. 
In particular, it 
models the coverage 
areas of wireless devices whose radio signals can penetrate 
up to $k$ walls~\cite{aichholzer2009modem,fabila2009modem}. 
This makes the problem particularly interesting for the 
limited workspace model, since these wireless devices are typically 
equipped with only a small amount of memory for computational tasks 
and may need to determine their coverage region using the 
few resources at their disposal.

The notion of $k$-visibility has previously been considered
in the context of art-gallery-style questions~\cite{ballinger2013coverage,eppstein2007guard,
fulek2009intersecting,
o2012computational} and in the definition
of certain geometric graphs~\cite{dean2005bar,
felsner2008parameters,hartke2007further}.
While the $0$-visibility region is always connected,
the $k$-visibility region may have several components.
Bajuelos~\etal~\cite{bajuelos2012hybrid} present an 
algorithm for a slightly different notion of 
$k$-visibility. It computes the region of the \emph{plane} 
which is $k$-visible from $q$ in the presence of a 
simple polygon $P$ with $n$ vertices, using $O(n^{2})$ 
time and $O(n^2)$ space. In this setting, the 
$k$-visibility region is connected. We believe that our 
ideas are also applicable for this notion and lead to an
improvement of their result.\footnote{The algorithm of 
Bajuelos~\etal~\cite{bajuelos2012hybrid} essentially first computes  
a complete arrangement of quadratic size that encodes the whole 
visibility information, and then extracts the $k$-visible region from 
this arrangement. Our algorithms, on the other hand, use a 
plane sweep so that only the relevant parts of this arrangement are 
considered. Thus, when $O(n)$ words of workspace are available, 
we achieve a running time of $O(n \log n)$.}

\paragraph{Related work.}
The optimal classic algorithm for computing the 
$0$-visibility region needs $O(n)$ time and $O(n)$ 
space~\cite{joe1987corrections}. In the 
\emph{constant-workspace model} (i.e., for $s = 1$), the
$0$-visibility region of a point $q \in P$ can
be reported in $O(n\bar{r})$ time, where 
$\bar{r}$ is the number of \emph{reflex} vertices 
of $P$ that occur in the output,
as shown by Barba~\etal~\cite{barba2014computing}.
This algorithm scans the boundary $\partial P$
in counterclockwise order, and it reports the 
maximal subchains of $\partial P$ that are 
$0$-visible from $q$. More precisely, this 
works as follows: we find a vertex
$v_\text{start}$ of $P$ that is $0$-visible 
from $q$. Walking from $v_\text{start}$, we 
then go until the next reflex vertex
$v_\text{vis}$ that is $0$-visible from $q$,
in counterclockwise direction.
This takes $O(n)$ time. The first intersection 
of the ray $qv_\text{vis}$ with $\partial P$ is 
called the \emph{shadow} of $v_\text{vis}$. 
Now, the end vertex of the maximal counterclockwise 
visible chain starting at $v_\text{start}$ is either 
$v_\text{vis}$ or its shadow. In each case, 
the next maximal visible chain starts at the other
of the two vertices ($v_\text{vis}$ or its shadow). Thus, we  can
find a maximal visible chain and a new starting 
point in $O(n)$ time. 
The number of iterations is $\bar{r}$, the number 
of reflex vertices that are $0$-visible from $q$.
This gives an algorithm with $O(n\bar{r})$ running 
time and $O(1)$ workspace. 

Now suppose that the number of reflex vertices in
$P$ with respect to $q$ is $r$. If the 
available workspace is $O(s)$, for 
$s \in \{1, \dots, O(\log r)\}$, Barba~\etal~\cite{barba2014computing} 
show how to find the $0$-visibility region 
of $q$ in $P$ in $O({nr}/{2^s}+n\log^2{r})$ 
deterministic time or $O({nr}/{2^s}+n\log{r})$ 
expected time. Their method is recursive. 
It uses the previous algorithm as the base,
and in each step of the recursion, it splits a 
chain on $\partial P$ into two subchains that each 
contains roughly half of the visible reflex vertices 
of the original chain. Since the $0$-visibility region 
and the $k$-visibility region of $q$ for $k > 0$ have 
different properties, there seems to be no 
straightforward way to generalize this approach
to our setting.
Later, Barba~\etal~\cite{barba2015space} provided
a general method for obtaining time-space trade-offs for
\emph{stack-based} algorithms. 
This gives an alternative trade-off for 
computing the $0$-visibility region:
there is an algorithm that 
runs in $O(n^2\log n/2^s)$ 
time for $s = o(\log n)$ and 
in $n^{1+O(1/\log s)}$ time for $s \geq \log n$.\footnote{The 
actual trade-off is more nuanced, 
but we simplified the bound to make it more digestible for the
casual reader.}
Again,
this approach does not seem to be directly applicable
to our setting.

Abrahamsen~\cite{Abrahamsen13} presents a constant workspace
algorithm that computes the visible part of one edge
from another edge in a simple polygon $P$ in $O(n)$ time,
where $n$ is the number of vertices in $P$.
This gives an algorithm that needs $O(mn)$ 
time and $O(1)$ words of workspace to compute
the weak visibility region of one edge in $P$.
The parameter $m$ denotes the size of the resulting
weak visibility polygon.

\paragraph{Our Results.}
We look at the more general problem of 
computing the $k$-visibility region of a simple 
polygon $P$ for a given point $q\in P$.
We give a constant workspace algorithm for this
problem, and we establish a time-space trade-off. 
Our first algorithm runs in $O(kn + cn)$ time
using $O(1)$ words of space, and our second 
algorithm requires $O(cn/s+c\log{s} + 
\min\{\lceil k/s \rceil n,n \log{\log_s{n}}\})$ 
expected time and $O(s)$ words of workspace. 
Here, $c \in \{1, \dots, n\}$ is the number of 
\emph{critical vertices} of $P$ for $q$, where the 
$k$-visibility region of $q$ may change. A precise
definition is given later. 

We generalize this result for polygons with holes 
and for sets of non-crossing line segments.
More precisely, we show that in a polygon $P$
with $h$ holes, we can report the $k$-visibility
region of a point $q \in P$ in expected time 
$O(cn/s+c\log{s} + 
\min\{\lceil k/s \rceil n,n \log{\log_s{n}}\})$
using $O(s)$ words of workspace. In an
arrangement of $n$ pairwise non-crossing line
segments, this takes $O(n^2/s + n\log s)$ 
deterministic time.

%-------------------------------------------------------------------------
\section{Preliminaries and Definitions}

\begin{figure}
 \centering 
 \includegraphics{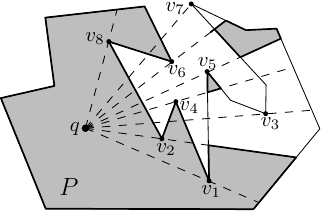}
\caption{An example with $k=2$. 
The hatched regions are not $2$-visible for $q$.
The vertices $v_1, \dots, v_8$ are critical for $q$.
More precisely, $v_1, v_2, v_3, v_6$ are 
start vertices, and $v_4, v_5, v_7, v_8$ 
are end vertices. $\partial P$ is 
partitioned into $8$ disjoint chains, e.g, 
the counterclockwise chain $v_3v_5$.}
\label{fig:fig1}
\end{figure}

Let $s \in \{1, \dots, n\}$ be the amount of available workspace,
measured in words.
We assume that the input polygon $P$ is given 
as a sequence of $n$ vertices in counterclockwise (CCW) 
order along $\partial P$.
The input also contains the query point $q\in P$ and
the visibility parameter $k \in \{0, \dots, n - 1\}$. The aim is to 
report $V_k(P,q)$, using $O(s)$ words 
of workspace. We require that the input is in \emph{weak 
general position}, i.e., the query point $q$ does not lie on any 
line through two distinct vertices of $P$. Without loss of generality, 
we assume that $k$ is even: if $k$ is odd, we 
can just compute $V_{k-1}(P,q)$, which is the same as
$V_k(P,q)$, by definition. The boundary $\partial V_k(P,q)$ of 
$V_k(P,q)$ consists of pieces of $\partial P$ and chords 
of $P$ that connect two such pieces; see Figure~\ref{fig:fig1}.

%%%%%%%%%%%%%%%%
%%%%%%%%%%%%%%%%%

We fix a coordinate system with origin $q$.
For $\theta \in [0, 2\pi)$, we denote by $r_\theta$ the ray 
that emanates from $q$ and has CCW-angle $\theta$ with the 
$x$-axis. An edge of $P$ that intersects $r_\theta$ is called 
an \emph{intersecting edge} of $r_\theta$. The \emph{edge list} 
of $r_\theta$ is defined as the list of intersecting edges of 
$r_\theta$, sorted according to their intersection with $r_\theta$,
in increasing distance from $q$. The $j^\text{th}$ element of this 
list is denoted by $e_{\theta}(j)$. We also say that $e_{\theta}(j)$ has
\emph{rank} $j$ in the edge list of $r_\theta$, or simply on $r_\theta$.

The \emph{angle} of a vertex $v$ of $P$
refers to the angle $\theta \in [0, 2\pi)$ at which 
$r_\theta$ encounters $v$. Suppose $r_\theta$ stabs a vertex $v$ of $P$.
We call $v$ a \emph{critical}
vertex if its incident edges lie on the same side of $r_\theta$, and
a \emph{non-critical} vertex otherwise. 
We can check in constant time whether a given vertex of $P$ is 
critical. We use $c$ to denote the number of 
critical vertices in $P$. Let $v$ be a critical 
vertex. We call $v$ a 
\emph{start vertex} if both incident edges lie counterclockwise
of $r_\theta$, and an \emph{end vertex} otherwise; see 
Figure~\ref{fig:fig1}.
A \emph{chain} is a sequence of edges of $P$ 
(in CW or CCW order along $\partial P$) which starts at a start vertex
and ends at an end vertex and contains no other
critical vertices. Note that every ray $r_\theta$ 
intersects each chain at most once. Thus, we will sometimes
talk of \emph{chains} that appear in the edge list of 
a ray $r_\theta$.

Suppose we continuously increase $\theta$ from $0$ to $2\pi$.
The edge list of $r_\theta$ only changes when $r_\theta$ 
encounters a vertex $v$ of $P$. This change only involves 
the two edges incident to $v$.
At a non-critical vertex $v$, the edge list is updated by
replacing one incident edge of $v$ with the other. The other edges and
their order in the edge list do not change.
At a critical vertex $v$, the edge list is updated by
adding or removing both incident edges of $v$, depending
on whether $v$ is a start vertex or an end vertex. The other edges 
and their order in the edge list are not affected; see 
Figure~\ref{fig:fig1}. 
If $r_\theta$ stabs a start vertex of $P$, we define the edge list 
of $r_\theta$ to be the edge list of $r_{\theta+\eps}$, for a 
small enough $\eps >0$. 
If $r_\theta$ stabs an end vertex or a non-critical vertex of $P$, 
we define the edge list 
of $r_\theta$ to be the edge list of $r_{\theta-\eps}$, for a 
small enough $\eps >0$.

For any $\theta \in [0, 2\pi)$,
only the first $k+1$ elements 
in the edge list of $r_\theta$ are $k$-visible from $q$ in
direction $\theta$. While increasing $\theta$, as
long as $r_\theta$ does not encounter a 
critical vertex, the $k$-visible chains in direction $\theta$ do not change.
However, if $r_\theta$ encounters a critical vertex $v$, then 
this may affect which chains are visible from $q$.
This happens if at least one of the incident edges to $v$ is among the 
first $k+1$ elements in the edge list of $r_\theta$. 
In other words, if $v$ is $k$-visible from $q$, which means that $v$ 
does not lie after $e_{\theta}(k+1)$ on $r_\theta$. 
The next lemma shows that in this case a segment on $r_\theta$ may 
occur on $\partial V_k(P,q)$. 

\begin{figure}
 \centering 
 \includegraphics{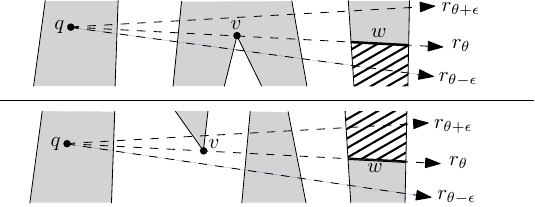}
\caption{An example with $k = 4$. 
The hatched regions are not 4-visible for $q$.
(a) The ray $r_\theta$ encounters the end vertex $v$.
The $4$-visibility region of $q$ right before $v$
extends to $e_\theta(5)$ and right after $v$ extends to $e_\theta(7)$.
(b) The ray $r_\theta$ encounters the start vertex $v$.
The $4$-visibility region of $q$ right before $v$
extends to $e_\theta(7)$ and right after $v$ extends to $e_\theta(5)$.
The segment $w$ in both figures is the window of $r_\theta$.}
\label{fig:fig2}
\end{figure}

\begin{lem}\label{lem:window}
Let $\theta \in [0, 2\pi)$ such that 
$r_{\theta}$ stabs a $k$-visible end or start vertex $v$. Then, the 
segment on $r_{\theta}$ between $e_{\theta}(k+2)$ and 
$e_{\theta}(k+3)$ is an edge of $V_k(P,q)$,
provided that these two edges exist.
\end{lem}
\begin{proof}
Suppose that $v$ is a $k$-visible end vertex. 
As mentioned above, right after $r_\theta$ encounters
$v$, two consecutive edges are removed from the
edge list of $r_\theta$.
Since $v$ is $k$-visible, these edges are among
the first $k+2$ entries in the edge list.
Thus, right after $v$, the $k$-visibility region of $q$ extends
to $e_\theta(k+3)$ (recall that the indices refer to
the situation just before $v$). Before $v$, the
$k$-visibility region extends to $e_\theta(k+1)$.
This means that the segment between $e_\theta(k+2)$ and $e_\theta(k+3)$
on $r_\theta$ belongs to 
$\partial V_k(P,q)$. In particular, this includes the case that
$e_\theta(k+1)$ and $e_\theta(k+2)$ are incident to $v$.
The situation for a $k$-visible start vertex $v$ is symmetric. 
Note that in this case, the indices in the edge list refer 
to the situation just after $v$; see Figure~\ref{fig:fig2}.
\end{proof}

%%%%%%%%%%%%%%
%%%%%%%%%%%%%%

Lemma~\ref{lem:window} leads to the following definition: 
let $\theta \in [0, 2\pi)$ such that
$r_\theta$ stabs a $k$-visible end or start
vertex $v$. The segment on $r_\theta$ between 
$e_{\theta}(k+2)$ and 
$e_{\theta}(k+3)$, if these edges exist, is called the \emph{window} 
of $r_{\theta}$; see Figure~\ref{fig:fig2}.

\begin{obs}\label{obs:complexity}
The $k$-visibility region $V_k(P,q)$ has $O(n)$ vertices.
\end{obs}
\begin{proof}
The boundary $\partial V_k(P,q)$ consists of subchains  
of $\partial P$ and of windows. Thus, a vertex of $V_k(P,q)$ is 
either a vertex of $P$ or an endpoint of a window. Since each 
critical vertex causes at most one window, since each
window has two endpoints, and since there are at most $n$ critical 
vertices, the total number of 
vertices of $V_k(P,q)$ is $O(n)$.
\end{proof}
%--------------------------------------------------------------------------
\section{A Constant-Memory Algorithm}\label{sec:constant-memory algorithm}

First, we assume that a constant amount of workspace is 
available. 
If the input polygon $P$ has no critical vertex, there 
is no window, and $V_k(P,q) = P$. This can be checked
in $O(n)$ time by a simple scan through the input.
Thus, we assume that 
$P$ has at least one critical vertex $v_0$. 
Again, $v_0$ can be found in $O(n)$ time with a single scan.
We choose our coordinate system such that
$q$ is the origin and such that $v_0$ lies on the positive $x$-axis.
We number the critical vertices of $P$ as $v_0, v_1,
\dots, v_{c-1}$ in the order that the ray $r_\theta$
encounters them. Let $\theta_i$ be the angle for $v_i$.
We simplify our notation and write $r_i$ instead
of $r_{\theta_i}$, and we let $e_{i}(j)$ denote the 
$j^\text{th}$ entry in the edge list of the ray $r_{i}$. 

We start with the ray $r_{0}$, and we find 
the edge $e_0(k+1)$ in $O(kn)$ time using $O(1)$ words of workspace. 
For this, we perform a simple \emph{selection} subroutine as follows:
we scan the input $k+1$ times, and in each pass, we find
the next intersecting edge of $r_{0}$ until 
$e_0(k+1)$. If $v_0$ is $k$-visible, i.e., if it is not after  
$e_0(k+1)$ on $r_{0}$, we report the window of 
$r_{0}$, as given by Lemma~\ref{lem:window} (if it exists). 
Since the window is defined by $e_0(k+2)$ and $e_0(k+3)$, 
it can be found in two more scans over the input.

\begin{figure}
 \centering 
 \includegraphics{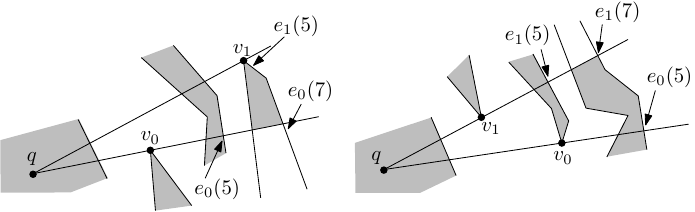}
\caption{Two cases for going from $v_0$ to $v_1$, with
$k = 4$. 
(a) Both $v_0$ and $v_1$ are end vertices. We
use $e_0(5)$ to find $e_0(7)$ and follow the chain until
$e_1(5)$.
(b) Both $v_0$ and $v_1$ are start vertices.
We follow the chain of $e_0(5)$ until $e_1(7)$, and
then use it to find $e_1(5)$.
We report the window from $e_1(6)$ to $e_1(7)$.}
\label{fig:fig3}
\end{figure}

Next, we find $v_1$ by a single scan of $\partial P$.
Then, we determine $e_1(k+1)$. 
This can be done in $O(n)$ time 
by using $e_0(k+1)$ as a starting point: we know that 
if $v_{0}$ is an end
vertex, the two incident chains of $v_0$ disappear
in the edge list of $r_1$. If $v_1$ is a start
vertex, the two incident chains of $v_1$ appear 
in the edge list of $r_1$.
All other chains are not affected, and they intersect $r_0$ and 
$r_1$ in the same order. 
Using this, we first find the edge $e'$ that has
rank $k+1$ in the edge list of the ray $r_{\theta_0 + \eps}$ just after
$r_0$. Depending on the type and position of $v_0$,
$e'$ is either $e_0(k+1)$ or $e_0(k+3)$, and it can
be found in $O(n)$ time. Then,
by scanning $\partial P$ starting from $e'$, we can
find the edge $e''$ on the chain of $e'$ that intersects
the ray $r_{\theta_1 - \eps}$ just before $r_1$, again
in $O(n)$ time. Depending on the type and position of $v_1$, 
the edge $e''$ is either $e_1(k+1)$ or $e_1(k+3)$. 
Thus, we can find $e_1(k+1)$ using $e''$ in $O(n)$ time; 
see Figure~\ref{fig:fig3}.

%%%%%%%%%%%%%%
%%%%%%%%%%%%%%

If $v_1$ is $k$-visible, we report the window of 
$r_{1}$ in $O(n)$ time, as described above.
Finally, we
report the subchains of $\partial V_k(P,q)$ between $r_{0}$ 
and $r_{1}$ by scanning $\partial P$. More precisely, 
we walk along $\partial P$ in counterclockwise direction.
Whenever we enter the counterclockwise cone between $r_0$ and $r_1$,
we check whether the intersection between $\partial P$ and
$r_0$ or $r_1$ occurs at or before $e_0(k+1)$ or $e_1(k+1)$, respectively.
If so, we report the subchain of $\partial P$ until
we leave the cone again.

We repeat this procedure until all critical vertices 
have been processed; see Algorithm~\ref{alg:pseudocode1}. 
Here and in the following algorithms, if there are 
less than $k+1$ intersecting edges on $r_i$, we store the 
last intersecting edge together with its rank.
We use this edge instead of
$e_{i}(k+1)$, in the procedure above, to find $e_{i+1}(k+1)$ or the
last intersecting edge of $r_{i+1}$ and its rank.
The number of 
critical vertices is $c$. For each of them, we spend
$O(n)$ time. Additionally, the selection subroutine for 
$v_0$ takes $O(kn)$ time. 
This leads to the following theorem:

%%%%%%%%%%%%%%%%%%%%%%%%%%%
\begin{alg}
 \KwIn{Simple polygon $P$, point $q\in P$, $k \in \N$}
 \KwOut{The boundary of the $k$-visibility region of $q$ in $P$,
 $\partial V_k(P,q)$}
 \If {$P$ has no critical vertex}{
    return $\partial P$\;
 }
 $v_0 \gets$ a critical vertex of $P$\;
 Find $e_0(k+1)$ using selection\;
 $i\gets 0$\; 
 \Repeat{$v_i = v_0$}{
 	\If {$v_i$ lies on or before $e_i(k+1)$ on $r_i$}{
 		Report the window of $r_i$ (if it exists)\;
 	}
 	$v_{i+1} \gets$ the next counterclockwise critical vertex after 
	$v_i$\;
 	Find $e_{i+1}(k+1)$ using $e_{i}(k+1)$\;
 	Report the part of $\partial V_k(P,q)$ between $r_i$ and $r_{i+1}$\;
 	$i \gets i+1$
 	}	
 \caption{The constant workspace algorithm for computing
$V_k(P,q)$}\label{alg:pseudocode1}
\end{alg}
%%%%%%%%%%%%%%%%%%%%%%%%%%%

\begin{thm}\label{thm:constant-memory}
Given a simple polygon $P$ with $n$ vertices, a 
point $q\in P$, and a parameter $k \in \{0, \dots, n - 1\}$, 
we can
report the $k$-visibility region of $q$ in $P$ in $O(kn+cn)$ time using $O(1)$ 
words of workspace, where $c$ is the number of critical vertices of $P$.
\end{thm}
%-------------------------------------------------------------------------
\section{Time-Space Trade-Offs}

In this section, we assume that we have $O(s)$ words of workspace 
at our disposal, and we show how to exploit this additional 
workspace to compute the $k$-visibility region faster. We describe 
two algorithms. The first algorithm is a little simpler, and it is meant to 
illustrate the main idea behind the trade-off. Our main contribution 
is in the second algorithm, which is more complicated but achieves 
a better running time. In the first algorithm, we process the 
vertices in angular order in contiguous batches of size $s$. In each 
iteration, we find the next batch of $s$ vertices, and using the 
edge list of the last processed vertex, we construct a data 
structure that is used to output the windows of the batch. Using the 
windows, we report $\partial V_k(P,q)$ between the first and the last 
ray of the batch.\footnote{We emphasize that $\partial V_k(P,q)$
is not necessarily reported in order, but we ensure that the union 
of the reported line segments constitutes the boundary of the
$k$-visibility region.
}
In the second algorithm, we improve the running time by skipping the 
non-critical vertices. Specifically, in each iteration, we find the 
next batch of $s$ adjacent critical vertices, and as before, we 
construct a data structure for finding the windows. We need a more 
involved approach in order to maintain this data structure. 
The next lemma shows how to obtain the contiguous batches of 
vertices in angular order efficiently.  The procedure is taken  
from the work of Chan and Chen~\cite{chan2007multi} (see the second 
paragraph in the proof of Theorem 2.1 in~\cite{chan2007multi}).

\begin{lem}\label{lem:s-smallest}
Suppose we are given a read-only array $A$ with $n$ pairwise distinct 
elements from a totally ordered universe and an element $x \in A$. For 
any given parameter $s \in \{1, \dots, n\}$, there is an 
algorithm that runs in $O(n)$ time and uses $O(s)$ words of workspace 
and that finds the set of the first $s$ elements in $A$ that follow 
$x$ in the sorted order. 
\end{lem}

\begin{proof}
Let $A_{> x}$ be the subsequence of $A$ that contains exactly the 
elements in $A$ that are larger than $x$.
The algorithm makes a single pass over $A_{>x}$ and processes the 
elements in batches.
In the first step, we insert the first $2s$ elements of $A_{>x}$ 
into our workspace (without sorting them). We select the median 
of these $2s$ elements using $O(s)$ time and space, and we remove 
the elements which are larger than the median. In the next step, we 
insert the next batch of $s$ elements from $A_{>x}$ into the 
workspace, and we again find the median of the resulting $2s$ 
elements and remove those elements that are larger than the median. 
We repeat the latter step until all the elements of $A_{>x}$ have 
been processed. Clearly, at the end of each step, the $s$ smallest 
elements of $A_{>x}$ that we have seen so far reside in memory. 
Since the number of steps is $O(n/s)$ and since each step needs 
$O(s)$ time, the running time of the 
algorithm is $O(n)$. By construction, it uses $O(s)$ words 
of workspace.
\end{proof}

\begin{lem}\label{lem:k-selection}
Suppose we are given a read-only array $A$ with $n$ elements from 
a totally ordered universe and a number $k \in \{1, \dots, n - 1\}$. 
For any given parameter $s \in \{1, \dots, n\}$, there is an 
algorithm that runs in $O\big(\lceil k/s\rceil n\big)$ time and uses 
$O(s)$ words of workspace and that finds the $k^\text{th}$ smallest 
element in $A$.
\end{lem}
\begin{proof}
We  again process the elements of $A$ in batches.
In the first step, we apply Lemma~\ref{lem:s-smallest} to find the 
first batch with the $s$ smallest elements in $A$ and to put it into 
our workspace. This needs $O(n)$ time and $O(s)$ words of workspace. 
If $k \leq s$, we select the $k^\text{th}$ smallest element in the 
workspace in $O(s)$ time; otherwise, we find the largest element $x$ 
in the workspace, and we apply Lemma~\ref{lem:s-smallest} to find 
the set of $s$ elements following $x$. In step $i$, we apply 
Lemma~\ref{lem:s-smallest} to find
the $i^\text{th}$ batch of $s$ elements 
in the sorted order of $A$ and to insert this set of elements into the workspace. 
If $k \leq i\cdot s$, we select the $(k-(i-1)s)^\text{th}$ smallest 
element in the workspace in $O(s)$ time and we output it; otherwise, 
we find the largest element in the workspace and we continue. 
The element being sought is in the $\lceil k/s\rceil^\text{th}$ 
batch. Therefore, we can find it in 
$O\big(\lceil k/s\rceil n\big)$ time using $O(s)$ words of workspace.
\end{proof}

In addition to the simple algorithm in Lemma~\ref{lem:k-selection}, 
there are several other results on selection in the read-only model; 
see Table~1 of \cite{chan2014selection}. In particular, there is a
$O(n\log{\log_s{n}})$ expected time randomized algorithm for 
selection using $O(s)$ words of workspace in the 
limited workspace model~\cite{chan2010comparison,munro1996selection}. 
Depending 
on $k$, $s$, and $n$, we will choose the latter algorithm or 
the algorithm that we presented in Lemma~\ref{lem:k-selection}. 
In conclusion, the running time of selection in the 
limited workspace model using $O(s)$ words of workspace, 
denoted by $T_\text{selection}$, is 
$O(\min \{\lceil k/s\rceil n, n\log{\log_s{n}}\})$ expected time. 

%------------------------------------------------------------------
\subsection{First Algorithm: Processing All the Vertices}\label{sec:algo1}

Let $v_0$ be some vertex of $P$. 
We choose our coordinate system such that
$q$ is the origin and such that $v_0$ lies on the positive $x$-axis.
We apply Lemma~\ref{lem:s-smallest} to find the batch of $s$ 
vertices with the smallest positive angles, and we 
sort them in workspace in $O(s \log s)$ time. Let $v_1, \dots, v_s$ 
denote these vertices in sorted order. We use the 
selection subroutine (with $O(s)$ words of workspace) to find 
$e_0(k+1)$ on $r_0$, and if $v_0$ is a $k$-visible vertex, i.e., if 
it does not occur after $e_0(k+1)$ on $r_0$, we report
its window (if it exists).  Recall that if there are less than $k+1$
intersecting edges on $r_0$, we store the last intersecting edge 
together with its rank.

Then, we apply Lemma~\ref{lem:s-smallest} four times in order to find 
the at most $4s + 1$ intersecting edges with ranks in 
$\{k - 2s + 1, \dots, k + 2s + 1\}$ on $r_0$
(Lemma~\ref{lem:s-smallest} can be applied, because we have 
$e_0(k+1)$ at hand). We insert these edges 
into a balanced binary search tree $T$, sorted according to 
their ranks on $r_0$. The edges in $T$ are candidates for having 
rank $k + 1$ on the next $s$ rays $r_1, \dots, r_s$. This is because,
as we explained in Section~\ref{sec:constant-memory algorithm}, 
if $e_i(k+1)$ belongs to the edge list of $r_{i-1}$, there is at 
most one edge between $e_{i-1}(k+1)$ and $e_i(k+1)$ in the edge 
list of $r_{i-1}$.
Therefore, if $e_i(k+1)$ appears in the edge list of $r_0$, 
there are at most $2i-1$ edges between $e_0(k+1)$ and
$e_i(k+1)$ in the edge list of $r_0$. 

\begin{figure}
 \centering 
 \includegraphics{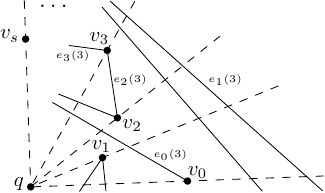}
\caption{The first batch $v_0, v_1, \dots, v_s$ of $s$ vertices 
  in angular order. The edge $e_1(3)$ is the second neighbor to the 
  right of $e_0(3)$ on $r_0$, because $v_0$ is an end vertex. 
  The edge $e_2(3)$ is the second neighbor to the left of $e_1(3)$ 
  which is inserted in $T$ before processing $v_2$. 
  The edge $e_2(3)$ is exchanged with $e_3(3)$, after processing 
  $v_3$, because $v_3$ is a non-critical vertex.}
\label{fig:fig4}
\end{figure}

Now the algorithm proceeds as follows: we go to the next vertex 
$v_1$, and we update $T$ depending on the types of $v_0$ and  
$v_1$: if $v_0$ is a non-critical vertex, we may need to exchange one 
incident edge of $v_0$ with another in $T$; if $v_0$ is an end vertex, 
we may need to remove its incident edges from $T$; and if 
$v_1$ is a start vertex, we may need to 
insert its incident edges into $T$.
In all other case, no action is necessary.
The insertion and/or deletion is performed only for the edges
whose ranks are between the smallest and the 
largest rank in $T$ (with respect to $r_1$). The update of $T$  
takes $O(\log s)$ time. Afterwards, we can find $e_1(k+1)$ and the 
window of $r_1$ (if it exists) in $O(1)$ time, using the position 
of $e_0(k+1)$ or its neighbors in $T$, as explained in 
Section~\ref{sec:constant-memory algorithm}.
See Figure~\ref{fig:fig4} for an example.

%%%%%%%%%%%%%%%%
%%%%%%%%%%%%%%%%

We repeat this procedure for $v_2, \dots, v_s$. We use, for 
$i = 2, \dots, s$, the binary search tree $T$ and the previous edge 
$e_{i-1}(k+1)$ in order to determine the next edge $e_i(k+1)$ and 
the window of $r_i$. This takes  $O(s \log s)$ total time. 
Whenever we find and report a window, we insert its endpoints into 
a balanced binary search tree $W$. This takes $O(\log s)$ time 
per window. The endpoints in $W$ are sorted according to 
their counterclockwise order along $\partial P$. For reporting 
the part of $\partial V_k(P, q)$ between $r_0$ and $r_s$, we use 
$W$ and the sequence $E = e_0(k + 1), e_1(k + 1), \dots, e_s(k + 1)$
of edges of rank $k+1$. 

For an edge $e$ of $P$, the 
\emph{$0s$-segment} of $e$ is the subsegment of $e$ that lies between
$r_0$ and $r_s$. If a $0s$-segment does not contain an endpoint 
of a window, then it is either completely $k$-visible or completely 
not $k$-visible. 
Thus, we can walk along $\partial P$ and, simultaneously, along the 
window endpoints in $W$. For each edge $e$ of $P$, we 
can check if the endpoints of the $0s$-segment of $e$ are $k$-visible
or not. We can do this in $O(1)$ time using $E$. With the help 
of the parallel traversal of $W$, we can also check 
if there is a window endpoint on $e$. This takes $O(|w_e|)$ 
time, where $|w_e|$ is the number of window endpoints on $e$. With 
this information, we can report the $k$-visible subsegments of 
the $0s$-segment of $e$. Since there are $O(n)$ window 
endpoints by Observation~\ref{obs:complexity},
and since we check each window endpoint once,
it follows that we need $O(n)$ time
to report the $k$-visible part of $\partial P$ between $r_0$ and 
$r_s$.

After processing $v_0, \dots, v_s$, we apply 
Lemma~\ref{lem:s-smallest} to find the next batch of $s$ 
vertices following $v_s$ in angular order. We sort them in 
$O(s\log s)$ time, using $O(s)$ words of workspace. The search tree 
$T$ for the previous
batch is not useful anymore, because it does not necessarily contain 
any right or left neighbor of $e_{s}(k+1)$ on $r_s$. Applying 
Lemma~\ref{lem:s-smallest} four times as before, we find the at most 
$4s + 1$ intersecting edges with ranks in $\{k - 2s + 1, \dots, k + 2s + 1\}$
on $r_s$, and we insert them into $T$. 
Then, as before, for each $s < i \leq 2s $, we find $e_i(k+1)$ and 
its corresponding window while maintaining $T$, $W$, and $E$. 
After that, we report the $k$-visible part of $\partial P$ between
$r_s$ and $r_{2s}$, where $r_{2s}$ is the ray for the last vertex 
in the batch, in sorted order. If $n$ is not divisible by $s$,
the last batch wraps around, taking the indices modulo $n$, 
but we report only the part of $\partial V_k(P,q)$ before $r_n = r_0$; 
see Algorithm~\ref{pseudocode2}. 

%%%%%%%%%%%%%%%%%%%%%%%%%%%
\begin{alg}
 \KwIn{Simple polygon $P$, point $q\in P$, $k \in \mathbb{N}$, $1 \leq s \leq n$}
 \KwOut{The boundary of $k$-visibility region of $q$ in $P$, 
   $\partial V_k(P,q)$}
 $v_0 \gets$ a vertex of $P$\;
 $E \gets  \langle e_0(k+1) \rangle$ (using the selection subroutine with $O(s)$ workspace)\;
 $T$, $W \gets$ an empty balanced binary search tree\;
 $i\gets 0$\;
 \Repeat{$i \geq n$}{
 	$v_{i+1}, \dots, v_{i + s} \gets$ sorted list of $s$ vertices following  $v_i$ in angular order \;
 	$T \gets \text{at most } 4s + 1 \text{ edges with rank in } \{k - 2s + 1,
    \dots, k+2s+1\} \text{ on } r_i$\;
 	\For {$j=i$ to $i + s - 1$}{
 		\If {$v_j$ lies on or before $e_j(k+1)$ on $r_j$}{
 			Report the window of $r_j$ (if it exists) \;
			Insert the endpoints of the window into $W$ (according to their 
              position on $\partial P$)\;
		}
		Update $T$ according to the types of $v_j$ and $v_{j+1}$\;
 		$E.\text{append}(e_{j+1}(k+1))$ (find it using $e_{j}(k+1)$ and $T$)\;
	}
	Report the part of $\partial V_k(P,q)$ between $r_i$ and $r_{\min\{i+s, n\}}$ (using $W$ and $E$)\;
 	$i \gets i + s$\;
 }	
 
 \caption{Computing $\partial V_k(P,q)$ using $O(s)$ words of workspace}\label{pseudocode2}
\end{alg}
%%%%%%%%%%%%%%%%%%%%%%%%%%%%

Overall, we need $O(n + s\log s)$ time for a batch. 
We repeat this procedure for $O(n/s)$ iterations, until all 
vertices are processed. 
Moreover, we run the selection subroutine in the first batch. 
Thus, the running time of the algorithm is 
$O(n/s(n+s\log{s}))+T_\text{selection}$. Since $T_\text{selection}$ 
is dominated by the other terms, we obtain the following theorem.

\begin{thm}\label{thm:limited-workspace}
Let $s \in \{1, \dots, n\}$.
Given a simple polygon $P$ with $n$ vertices in a read-only array, 
a point $q \in P$ and a parameter $k \in \{0, \dots, n - 1\}$, we
can report the $k$-visibility region of $q$ in $P$ in 
$O(n^2/s + n\log{s})$ time using $O(s)$ words of workspace.
\end{thm}
%----------------------------------------------------------------
\subsection{Second Algorithm: Processing only the Critical Vertices}
\label{algo2}

As in Section~\ref{sec:algo1}, we process the vertices in 
batches, but now we focus only on the critical vertices.
The new algorithm is similar to the algorithm in
Section~\ref{sec:algo1}, but it handles 
the data structure for the intersecting edges differently. In each 
iteration, we find the next batch of $s$ \emph{critical} vertices,
and we sort them in $O(s \log s)$ time using $O(s)$ words of 
workspace. As in the previous algorithm, we construct a data structure 
$T$ that contains the possible candidates for the edges of rank 
$k + 1$ on the rays for the $s$ critical vertices of the batch. 
In each step, we process the next critical vertex. We use $T$ 
to find the corresponding window, and we update $T$.
For updating $T$, we consider only the changes in the edge list
that are caused by the critical vertices. This is because
the non-critical vertices do not change the chains that appear in the 
edge list of the ray; they only affect the actual edge that intersects
it.\footnote{The algorithm in the published version of this
article is slightly different and relies on an
additional data structure $T_\text{aux}$ to update $T$.
Unfortunately, our running time analysis of this update strategy
was not correct. To fix this, we changed the update strategy
to the lazy method described here.}

More precisely, we use a \emph{lazy} strategy for updating $T$:
instead of always maintaining the edges that intersect the current ray, we
only store some edge on their corresponding chains, and we determine
the precise intersecting edges only when the need arises; see below, 
and Figure~\ref{fig:fig5} for an illustration.
After finding all the windows of the batch, we report 
the $k$-visible part of $\partial P$ between the first and the last 
ray of the batch.

As in Section~\ref{sec:constant-memory algorithm}, if $P$ has 
no critical vertex, then $V_k(P,q) = P$. This can be checked
in $O(n)$ time by a simple scan through the input.
Thus, we let $v_0$ be some critical vertex, and 
we choose our coordinate system such that
$q$ is the origin and such that $v_0$ lies on the positive $x$-axis.
In the first iteration, we compute $v_{1}, \dots ,v_{s}$, 
the list of $s$ critical vertices after $v_0$, sorted in angular 
order. Using Lemma~\ref{lem:s-smallest} and a traditional sorting 
algorithm, this takes $O(n + s \log s)$ time and $O(s)$ words of 
workspace.

Then, we process one critical vertex in each step.
In step~$0$, we find $e_0(k + 1)$ using our selection subroutine, and 
the at most $4s + 1$ intersecting edges with rank in 
$\{k - 2s + 1, \dots, k + 2s + 1\}$ on $r_0$. We insert them into a balanced 
binary search tree $T$, ordered according to their rank on $r_0$. 
This takes $T_\text{selection} + O(n + s\log s)$ time.
We use $e_0(k + 1)$ and $T$ to find and report the window of $r_0$ 
(if it exists).

%%%%%%%%%%%%%%%%
\begin{figure}
 \centering 
 \includegraphics{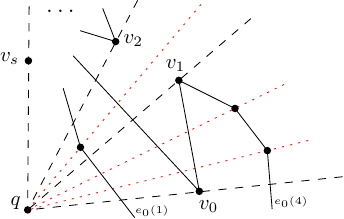}
\caption{The first batch $v_0, v_1, \dots, v_s$ of $s$ critical vertices 
  in angular order. The edge $e_0(1)$ intersects only $r_0$ and $r_1$.
  However, it is in the edge list of $r_2$ as a representative for $e_2(1)$.
  Similarly, $e_0(4)$ is a representative for $e_1(2)$ in the edge list 
  of $r_1$.}
\label{fig:fig5}
\end{figure}
%%%%%%%%%%%%%%%

In step~$1$, 
we update $T$ according to the types of $v_0$ 
and  $v_1$, so that $T$ contains representatives for 
the chains that intersect $r_1$: if 
$v_0$ is an end vertex, and if its incident edges are in $T$,
we remove those edges from $T$;
if $v_1$ is a start vertex, we insert the two incident edges
of $v_1$ 
(as representatives of the corresponding chains)
into $T$, provided that their ranks on $r_1$ 
are in the correct rank interval for the edges in $T$.
For finding the rank of the incident edges of $v_1$ on $r_1$, we 
perform a 
search in $T$ to compare the positions of these edges and the
elements of $T$ on $r_1$.
Whenever a comparison needs to be 
done with an edge $e$ stored in $T$, we check whether $e$
intersects $r_1$. If not, we follow the corresponding chain of $e$
until we find such an edge.
Thus, it takes $O(\log s + n_1')$ time to update $T$, where
$n_1'$ denotes the number of non-critical vertices that are traversed
to find the correct edges for comparisons during the update
operations. 

Now, $T$ contains 
at most $4s + 1$ intersecting chains of $r_1$. To determine  
$e_{1}(k+1)$, we walk along the chain of either $e_{0}(k+1)$ 
or its neighbors in $T$, until we meet the edge that 
intersects $r_1$.
Having $e_{1}(k+1)$ and $T$, we find and report
the window of $r_1$ (if it exists), again using our lazy strategy.
Finding $e_{1}(k+1)$ and the window of $r_1$
takes $O(n_1'')$ time, where $n_1''$ is the number of non-critical
vertices that are traversed during the search. 

In step $i = \{2, \dots, c-1 \}$, we repeat the same procedure
as in step $1$. We update $T$ for the edges that are incident
to the critical vertices $v_{i-1}$ 
and $v_i$. The only difference is that, if $v_{i-1}$ is an end vertex,
checking whether its chains are in $T$, and identifying
their representative edge in $T$, are not as straightforward as in step~$1$.
The problem is that we do not know which edge of each
chain has been stored in $T$ as its representative. To resolve this problem,
for any chain $C$ in $T$, we additionally store another edge of $C$
that is called the \emph{guide edge} and that is defined as follows:
if $C$ intersects $r_0$, the edge on $C$ that intersects $r_0$
is a \emph{type~$1$} guide edge;
and if $C$ has been inserted into $T$ in one of the
steps~$2, \dots, i-1$, the first edge of $C$ is a \emph{type~$2$} guide edge.

We store the type~$1$ guide edges in an array $G_1$, sorted according
to their rank on $r_0$, i.e., we copy the sorted elements of $T$ in step~$0$ 
into $G_1$.
The type~$2$ guide edges are stored in another array $G_2$,
sorted according to the step in which they have been inserted into $T$, 
i.e., the angle of the start vertex of the chain.
Therefore, in each step that new edges are inserted into $T$, 
those edges will also be added to $G_2$ in $O(1)$ time.
The keys for the elements in $G_2$ will be the corresponding rays
$r_i$. (since
both arrays have length $O(s)$, we can reserve memory for them
in advance.
The elements in $G_1$ and $G_2$ have 
cross-pointers to their corresponding entries in $T$.

To find the corresponding elements of the chains of an end vertex
$v_{i-1}$ in $T$, we walk backward on each chain, until
we either encounter an edge that intersects $r_0$ (a type~1 guide edge),
or the first edge of that chain (a type 2~guide edge). 
Then, by a binary search in $G_1$ or $G_2$ in $O(\log s)$ time,
and using the cross-pointer to the elements in $T$, we identify 
the corresponding entries in $T$.
Therefore, we can remove them from $T$ in $O(\log s)$ time.
Now, $T$ contains the chain list of $v_i$ and can be used
to find $e_{i}(k+1)$ with the help of $e_{i-1}(k+1)$. Finally, we
report the window of $r_i$ (if it exists), as in step $1$.
In total, processing the changes in $T$ for a batch 
takes $O(s\log s + n')$ time, where 
$n'$ is the number of non-critical vertices that lie between 
$r_0$ and $r_s$.

While processing the batch, we insert all $e_i(k+1)$, 
$0 \leq i \leq s$, into $E$. Also, whenever we find and report 
a window, we insert its endpoints, sorted according to their 
counterclockwise order along $\partial P$, into a balanced binary search 
tree $W$, in $O(\log s)$ time. After processing all the vertices 
of the batch, we use $W$ and $E$ to report the part of 
$\partial V_k(P, q)$ between $r_0$ and $r_s$, as in 
Section~\ref{sec:algo1}. The only difference is that now we 
keep track of the visibility of the whole chains between $r_0$ and $r_s$
instead of individual edges. As before, this takes $O(n)$ time.

In the subsequent iteration, we repeat the same procedure for the next 
batch of $s$ critical vertices. We repeat until all critical 
vertices are processed; see Algorithm~\ref{alg:pseudocode3}.
By construction, each non-critical vertex 
is handled in exactly one iteration. Since there 
are $O(c/s)$ iterations, updating $T$ 
takes $O(c\log s + n)$ time in total. All together, we get a 
total running time of $O(cn/s + c\log s)$, in addition to 
$T_\text{selection}$ in the first batch. This leads to the 
following theorem:

\begin{thm}\label{thm:improved-limited-workspace}
Let $s \in \{1, \dots, n\}$.
Given a simple polygon $P$ with $n$ vertices in a read-only array, 
a point $q \in P$ and a parameter $k \in \{0, \dots, n - 1\}$, we can 
report the $k$-visibility region of $q$ in $P$ in 
$O(cn/s+c\log{s}+\min\{\lceil k/s \rceil n,n\log{\log_s{n}}\})$ 
expected time using $O(s)$ words of workspace, 
where $c$ is the number of critical vertices of $P$ for $q$.
\end{thm}

%%%%%%%%%%%%%%%%%%%%%%%%%%%%
\begin{alg}
 \KwIn{Simple polygon $P$, point $q\in P$, $k \in \mathbb{N}$, $1\leq s\leq n$}
 \KwOut{The boundary of $k$-visibility region of $q$ in $P$, $\partial V_k(P,q)$}
 $v_0 \gets$ a critical vertex of $P$\;
 $E \gets  \langle e_0(k+1) \rangle$ (using the selection subroutine with $O(s)$ workspace)\;
 $T$, $W \gets$ an empty balanced binary search tree\; 
 $G_1, G_2 \gets \langle \rangle$\;
 $i\gets 0$\;
 \Repeat{$i \geq n$}{
 	$v_{i+1}, \dots, v_{i + s} \gets$ sorted list of $s$ critical 
	vertices following  $v_i$ in angular order \;
  	$T \gets \text{at most } 4s + 1 \text{ edges with rank in } 
	\{k - 2s + 1,
    \dots, k+2s+1\} \text{ on } r_i$\;
    $G_1 \gets$ the sorted elements of $T$\;
 	\For {$j=i$ to $i + s - 1$}{
 		\If {$v_j$ lies on or before $e_j(k+1)$ on $r_j$}{
 			Report the window of $r_j$ (if it exists) \;
			Insert the endpoints of the window into $W$ (sorted by position on $\partial P$)\;
		}
		\If {$v_j$ is an end vertex}{
			Find the guide edges of chains of $v_j$ (by walking along the chains)\;
			\If{the guide edges exist in $G_1$ or $G_2$}{
				Use their cross-pointers to find the corresponding elements in $T$\;
			Remove those corresponding elements from $T$
			}
		}
		\If {$v_{j+1}$ is a start vertex}{
			\Repeat{rank of the chains of $v_{j+1}$ on $r_{j+1}$ are determined}{
				Take the next edge $e$ along the 
				search path in $T$\;
				\If {$e$ does not intersect $r_{j+1}$}{
					Walk along the chain of $e$ until
					the edge $e'$ that intersects $r_{j+1}$\;
					Exchange $e$ with $e'$ in $T$\;
				}
				Compare the position of $e$ with $v_{j+1}$ on $r_{j+1}$\;
			}
			\If {the rank of the chains of $v_{j+1}$ are valid for $T$}{
				Insert the chains of $v_{j+1}$ into $T$ according to their rank on $r_{j+1}$\;
				Append the chains of $v_{j+1}$ to $G_2$\;
			}
		}
	Find $e_{j+1}(k+1)$ by walking along $e_{j}(k+1)$ or its neighbors in $T$\;
	Append $e_{j+1}(k+1)$ to $E$\;	
	}
	Report subchains of $\partial V_k(P,q)$ between $r_i$ and $r_{\min\{i+s, n\}}$ (using $W$ and $E$)\;
 	$i \gets i + s$\;
 }	 
 \caption{Computing $\partial V_k(P,q)$ using $O(s)$ words of workspace}\label{alg:pseudocode3}
\end{alg}
%%%%%%%%%%%%%%%%%
%--------------------------------------------------------------------
\section{Variants and Extensions}

Our results can be extended in several ways; for example, computing 
the $k$-visibility region of a point $q$ inside a polygon $P$, where 
$P$ may have holes, or computing the $k$-visibility region of a 
point $q$ in a planar arrangement of $n$ non-crossing segments 
inside a bounding box (the bounding box is only for bounding the 
$k$-visibility region). Concerning the first extension, all the properties
we showed to hold for the algorithms for simple polygons
also hold for the case with holes. The only noteworthy 
issue is the use of $\partial P$ to 
report the $k$-visible segments of $\partial P$. In the case of 
polygons with holes, after walking on the outer part of 
$\partial P$, we walk on the boundaries of the holes one by one and 
we apply the same procedures for them. If there is no window on the 
boundary of a hole, then it is either completely $k$-visible or 
completely non-$k$-visible. For such a hole, we check if it is 
$k$-visible and, if so, we report it completely. 
This leads to the following corollary:

\begin{cor}\label{cor:polygon with holes}
Let $s \in \{1, \dots, n\}$.
Given a polygon $P$ with $h \geq 0$ holes and $n$ vertices in a 
read-only array, a point $q \in P$ and a parameter 
$k \in \{0, \dots, n - 1\}$, we can report the $k$-visibility region of $q$ in $P$ in  
$O(cn/s+c\log{s}+\min\{\lceil k/s\rceil n,n\log{\log_s{n}}\})$ 
expected time using $O(s)$ words of workspace. Here, $c$ is the 
number of critical vertices of $P$ for the point $q$.
\end{cor}

Concerning the second problem, for a planar arrangement of $n$ non-crossing 
segments inside a bounding box, the output consists of the 
$k$-visible parts of the segments. All the segments endpoints 
are critical vertices and should be processed. In the parts of the 
algorithm where a walk on the boundary is needed, a sequential scan 
of the input leads to similar results. Similarly, there may be some 
segments with no window endpoints. For these, we only need to check 
visibility of an endpoint to decide whether they are completely 
$k$-visible or completely non-$k$-visible. 
This leads to the following corollary:

\begin{cor}\label{cor:segments}
Let $s \in \{1, \dots, n\}$.
Given a set $S$ of $n$ non-crossing planar segments in a read-only 
array that lie in a bounding box $B$, a point $q \in B$ and a 
parameter $k \in \{0, \dots, n - 1\}$, there is an algorithm that 
reports the $k$-visible subsets of segments in $S$ from $q$ in
$O(n^2/s+n\log{s})$ time using $O(s)$ words of workspace.
\end{cor}

\section{Conclusion}
We have proposed algorithms for a class of $k$-visibility problems 
in the limited workspace model, and we have provided time-space 
trade-offs for these problems. We leave it as an open question 
whether the presented algorithms are optimal. Also, it 
would be interesting to see whether there exists an output sensitive 
algorithm whose running time depends on the number of windows in 
the $k$-visibility region, instead of the critical vertices in the input polygon.

Finally, our ideas are also applicable to the slightly
different definition of $k$-visibility used by Bajuelos et 
al.~\cite{bajuelos2012hybrid}. Thus, our techniques can be used to 
improve their result, achieving $O(n \log n)$ running time if 
$O(n)$ words of workspace are available.

%-------------------------------------------------------------------
% Bibliography
\bibliographystyle{abbrv}
\bibliography{ref}

\end{document}